\documentclass[fleqn,usenatbib]{mnras}

\usepackage{newtxtext,newtxmath}

\usepackage{hyperref}

\usepackage[T1]{fontenc}

\DeclareRobustCommand{\VAN}[3]{#2}
\let\VANthebibliography\thebibliography
\def\thebibliography{\DeclareRobustCommand{\VAN}[3]{##3}\VANthebibliography}

\def\mearth{{\rm\,M_\oplus}}

\usepackage{graphicx}	
\usepackage{amsmath}	

\title[Oort cloud (exo)planets]{Oort cloud (exo)planets}

\author[Raymond, Izidoro, \& Kaib]{Sean N. Raymond$^1$\thanks{E-mail: rayray.sean@gmail.com}, 
Andre Izidoro$^{2,3}$, and 
Nathan A. Kaib$^{4,5}$
\\
$^\mathrm{1}$Laboratoire d'Astrophysique de Bordeaux, CNRS and Universit{\'e} de Bordeaux, All{\'e}e Geoffroy St. Hilaire, 33165 Pessac, France \\
$^2$Department of Physics and Astronomy, 6100 Main MS-550, Rice University, Houston, TX 77005, USA \\
$^3$Department of Earth, Environmental and Planetary Sciences, MS 126, Rice University, Houston, TX 77005, USA\\
$^4$HL Dodge Department of Physics and Astronomy, University of Oklahoma, Norman, OK 73019, USA \\
$^5$Planetary Science Institute, 1700 E. Fort Lowell, Suite 106, Tucson, AZ 85719, USA
}

\date{Accepted XXX. Received YYY; in original form ZZZ}

\pubyear{2022}

\begin{document}
\label{firstpage}
\pagerange{\pageref{firstpage}--\pageref{lastpage}}
\maketitle

\begin{abstract}
Dynamical instabilities among giant planets are thought to be nearly ubiquitous, and culminate in the ejection of one or more planets into interstellar space. Here we perform N-body simulations of dynamical instabilities while accounting for torques from the galactic tidal field.  We find that a fraction of planets that would otherwise have been ejected are instead trapped on very wide orbits analogous to those of Oort cloud comets.  The fraction of ejected planets that are trapped ranges from 1-10\%, depending on the initial planetary mass distribution.  The local galactic density has a modest effect on the trapping efficiency and the orbital radii of trapped planets. The majority of Oort cloud planets survive for Gyr timescales. Taking into account the demographics of exoplanets, we estimate that one in every 200-3000 stars could host an Oort cloud planet.  This value is likely an overestimate, as we do not account for instabilities that take place at early enough times to be affected by their host stars' birth cluster, or planet stripping from passing stars.  If the Solar System's dynamical instability happened after birth cluster dissolution, there is a $\sim$7\% chance that an ice giant was captured in the Sun's Oort cloud.

\end{abstract}




\section{Introduction}

The 'planet-planet scattering' model is the leading hypothesis to explain the broad eccentricity distribution of giant exoplanets~\citep{butler06,udry07}.  The model proposes that systems of gas giants undergo dynamical instabilities that frequently eject planets into interstellar space; the eccentric orbits of surviving planets are scars from these violent events~\citep{rasio96,weidenschilling96,lin97,adams03,chatterjee08,juric08,raymond10}.  This model can also explain other observed features of giant exoplanet systems, such as their dynamical spacing and secular architectures~\citep[e.g.][]{raymond09a,timpe13,bitsch23}.  

The Solar System is also thought to have undergone a dynamical instability early in its history, albeit a relatively mild one~\citep{tsiganis05,morby07}. There is abundant (circumstantial) evidence to support this, from the  distributions of small body populations and the giant planets' orbital architectures~\citep[see][]{nesvorny18b}. During the instability, one or two extra ice giants may have been ejected~\citep{nesvorny12,batygin12b}.

Previous work on dynamical instabilities has rarely taken into account the external Galactic environment. Stars are born in clusters~\citep[e.g.][]{adams10}, and after clusters disperse stars remain in loose dynamical contact with other stars while orbiting within the Galaxy's gravitational field~\citep{oort50}.  It is vital to account for the galactic tidal field (specifically the component that results from the density gradient perpendicular to the galactic disk) to model the origin and dynamics of Oort cloud comets, a fraction of which are trapped on wide orbits due to torques from the Galactic tide or from passing stars~\citep{heisler86,duncan87,wiegert99,kaib08}. These external factors also affect the fates of scattered planets.  

Here we show that many planets that would have been considered `ejected' in previous work are actually trapped on Oort cloud-like orbits with orbital radii of tens of thousands of au. We quantify the importance of the Galactic tidal field and estimate the frequency of Oort cloud exoplanets.  

\section{Simulations}

\subsection{Code and Initial conditions}

Our simulations capture the dynamics of unstable planetary systems while taking into account the Galactic tidal field. We modified the {\tt Mercury} integrator~\citep{chambers99} to include the galactic tide, using the same methodology as previous studies~\citep{heisler86,wiegert99,levison01b,kaib08}.  The calculation is done in Cartesian galactic coordinates ($\Tilde{x},\Tilde{y},\Tilde{z}$), where $\Tilde{x}$ points away from the Galactic center, $\Tilde{y}$ points in the direction of Galactic rotation, and $\Tilde{z}$ points toward the Galactic South pole.  The resulting force can be expressed as:
\begin{equation}
    F_{tide} = \left(\mathrm{A}-\mathrm{B}\right) \left(3~\mathrm{A}+\mathrm{B}\right) \Tilde{x} \hat{\Tilde{x}}-
    \left(\mathrm{A}-\mathrm{B}\right)^2 \Tilde{y} \hat{\Tilde{y}}-
    \left(4\pi G \rho_0 - 2*(\mathrm{B}^2-\mathrm{A}^2)\right)\Tilde{z} \hat{\Tilde{z}},
\end{equation}
\noindent where $\mathrm{A} = 14.82$~km/s/kpc and $\mathrm{B} = 12.37$~km/s/kpc are Oort's constants, and $\rho_0$ is the local Galactic density, which we set to $0.15~\mathrm{M_\odot} \ pc^{-3}$ for our fiducial set of simulations~\citep[following][]{tremaine93}, and test the effect of varying   $\rho_0$ in Section 2.4.  We assumed a constant angle of $60.2^\circ$ between the ecliptic and Galactic planes, following \cite{kaib08}. The galactic distribution of inclinations should be uniform in $cos(i)$, such that the Solar System value is almost exactly the median.  The galactic torque scales as $sin(i)^2$, so systems with larger inclinations will experience modestly stronger torquing.  In practice, the galactic torque will affect all systems to some degree, even those in which the planets' initial plane is aligned with the galactic disk, because planet-planet scattering itself generates significant mutual inclinations~\citep[e.g.][]{chatterjee08,juric08}. 

We performed four sets of simulations with different planet masses, chosen because previous work demonstrated that they are compatible with the known giant exoplanet population~\citep{raymond08b,raymond09a,raymond10,raymond11,raymond12}.  The mass distributions are as follows:
\begin{itemize}
    \item The {\em mixed1} set: each system contains three planets with masses drawn from a $dN/dM \propto M^{-1.1}$ distribution~\citep{butler06,udry07} between 1 Saturn-mass and 3 Jupiter masses.  
    \item The {\em mixed2} set: each system started with four planets drawn from a $dN/dM \propto M^{-1.1}$ distribution between $10 \mearth$ and $3 M_{\mathrm{Jup}}$.
    \item The {\em equal-mass} set: three equal-mass planets of 1 Saturn-mass, 1 Jupiter-mass, and 4 Jupiter masses.
    \item The `Solar System-like' set mimics the Solar System's giant planet instability. Each simulation started with Jupiter, Saturn, and three ice giants~\citep[assuming one was ejected; see][]{nesvorny12} and an outer disk containing 270 planetesimal particles (that interacted gravitationally with the planets but not with each other) totalling $20 \mearth$.  
\end{itemize}

Our systems were designed to be immediately unstable but start from a plausible orbital configuration. For the {\em mixed1}, {\em mixed2} and {\em equal-mass} sets, the innermost planet was placed at 5 au. The orbital semimajor axis of each subsequent planet was randomly chosen to correspond to the location of the 3:4, 3:5, 2:3, or 1:2 with its inner neighbor (although other orbital angles were not chosen to favor resonance).  The orbital eccentricities of the inner and outer planets were randomly chosen below 0.01, but the eccentricity of the middle planet (or, for the {\em mixed2} set, either the second or third planet) was chosen that at periastron it crossed the orbit of the inner planet, ensuring an immediate dynamical instability.  For the Solar System-like set, the giant planets were placed in a multi-resonant configuration that represents a plausibly pre-instability configuration~\citep{nesvorny12,pierens14,clement21} with Jupiter at 5.35 au, Saturn in 2:3 resonance at 7.496 au, an additional $15 \mearth$ ice giant in 2:3 resonance with Saturn at 9.9~au, Uranus in 1:2 resonance with the extra ice giant at 15.7~au, and Neptune in 2:3 resonance with Uranus at 20.5~au. Unlike the other sets of simulations, the planets started on near-circular orbits. The starting inclinations of all planets were randomly chosen between zero and $1^\circ$. 

Each simulation was integrated for 100-200 Myr using a modified version of {\tt Mercury}'s hybrid integrator~\citep{chambers99}. 
We increased the size of the central (Solar-mass) star to 0.4 au because tests showed that our chosen timestep of 10 days accurately resolves approaches as close as this without accumulating errors in energy~\citep[see Appendix A in][]{raymond11}.  A planet was considered ejected if it reached the Sun's `tidal radius' of 170,000 au, which scales with the Galactic density and therefore the strength of the Galactic tide~\citep[][see Section 2.4]{tremaine93,wyatt17}.

\subsection{An example simulation}

\begin{figure}
\includegraphics[width=\columnwidth]{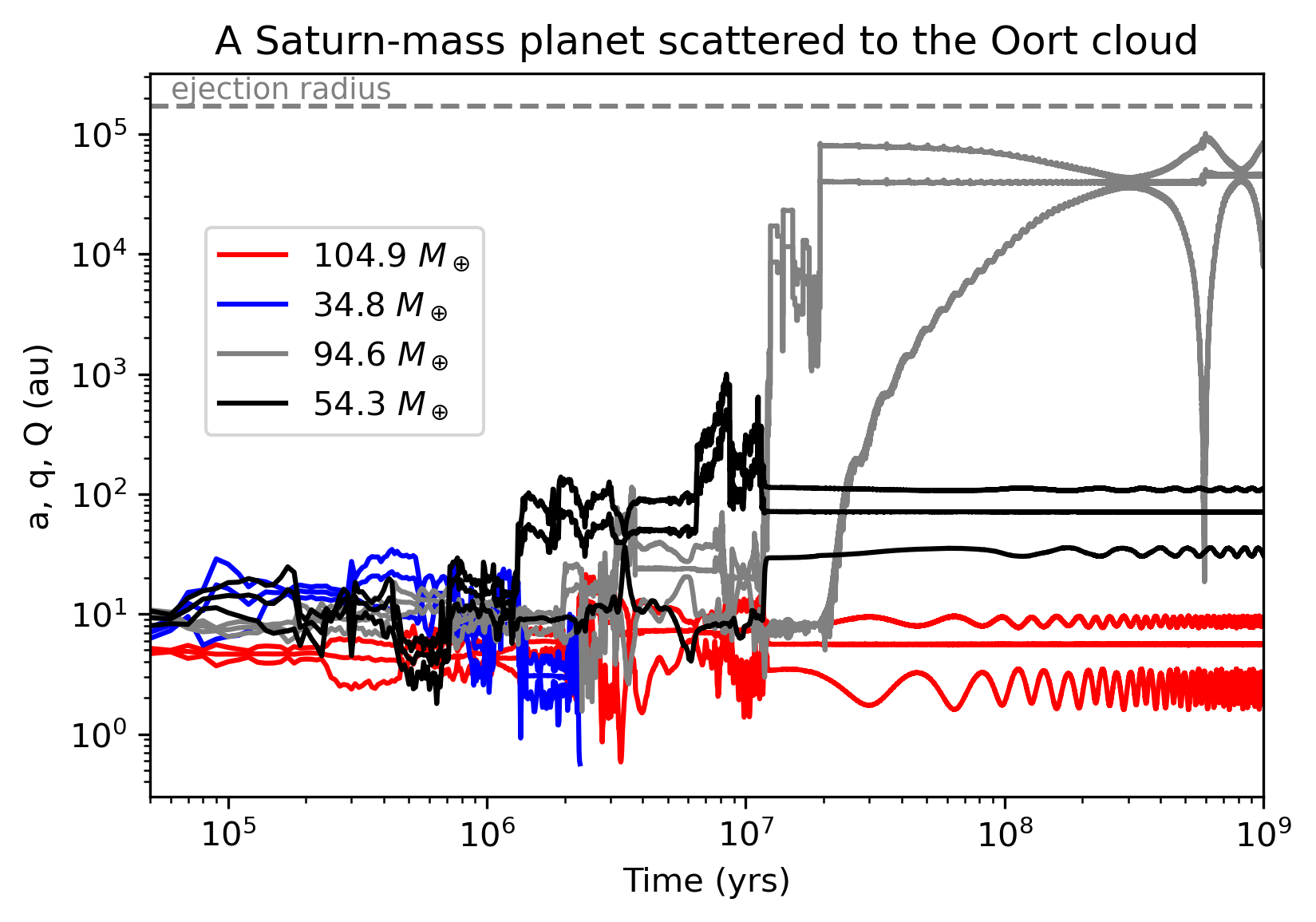}
    \caption{An example simulation from the {\em mixed2} set in which a roughly Saturn-mass planet was trapped on a very wide orbit.  Each planet's semimajor axis $a$, perihelion distance $q$, and aphelion distance $Q$ are shown. The dashed line shows the ejection radius at 170,000 au.}
    \label{fig:example}
\end{figure}

Figure~\ref{fig:example} shows a planet being trapped in its star's Oort cloud. This system (from the {\em mixed2} set) started with four planets, with masses -- in order of increasing orbital distance-- of $104.9 \mearth$, $34.8 \mearth$, $94.6 \mearth$, and $54.3 \mearth$. The dynamical instability led to close encounters between the planets, scattering each other to higher eccentricities such that the lowest-mass planet collides with the central star at $t \sim 2.3$~Myr.  At $\sim 10$~Myr, the $94.6 \mearth$ planet was repeatedly scattered by the more massive inner planet to a semimajor axis of $\sim 40,000$~au.  During the same phase of instability, the $54.3 \mearth$ was scattered out to an orbital radius close to 100~au. Over a span of $\sim 100$~Myr, the Galactic tide acted to decrease the $94.6 \mearth$ planet's orbital eccentricity, lifting its periastron distance and decoupling it from perturbations from the other planets.  At the end of the simulation, the three surviving planets had semimajor axes of 5.6, 71, and 39,153 au, eccentricities of 0.62, 0.53 and 0.27, and inclinations with respect to their starting plane of $65.7^\circ$, $8.6^\circ$ and $71.7^\circ$, respectively.  The two inner planets continued to undergo Lidov-Kozai-like oscillations in eccentricity and inclination on timescales of a few tens of Myr, but the outer planet was mostly decoupled.


\subsection{Results from the ensemble of simulations}

\begin{figure*}
\includegraphics[width=\columnwidth]{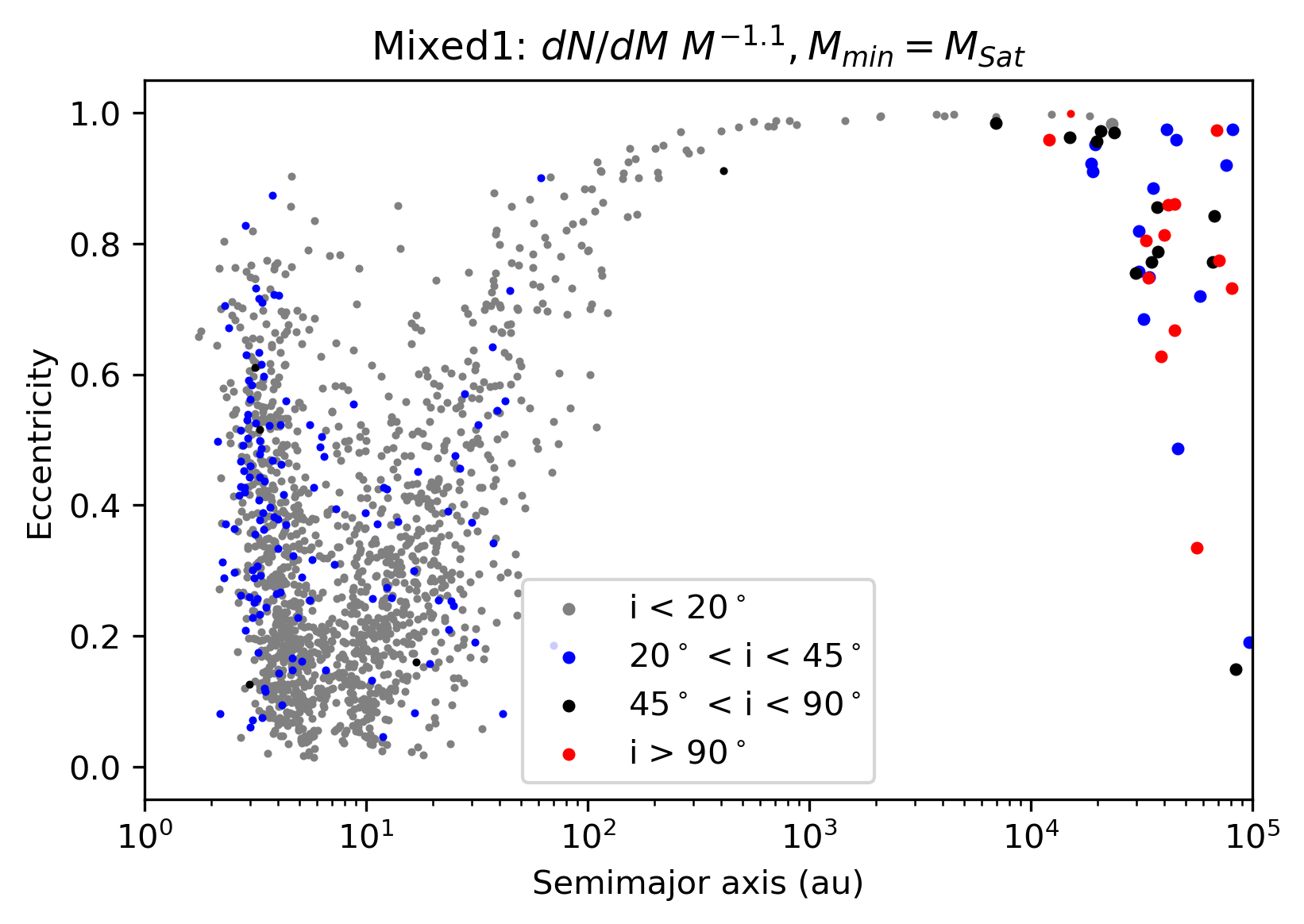}
\includegraphics[width=\columnwidth]{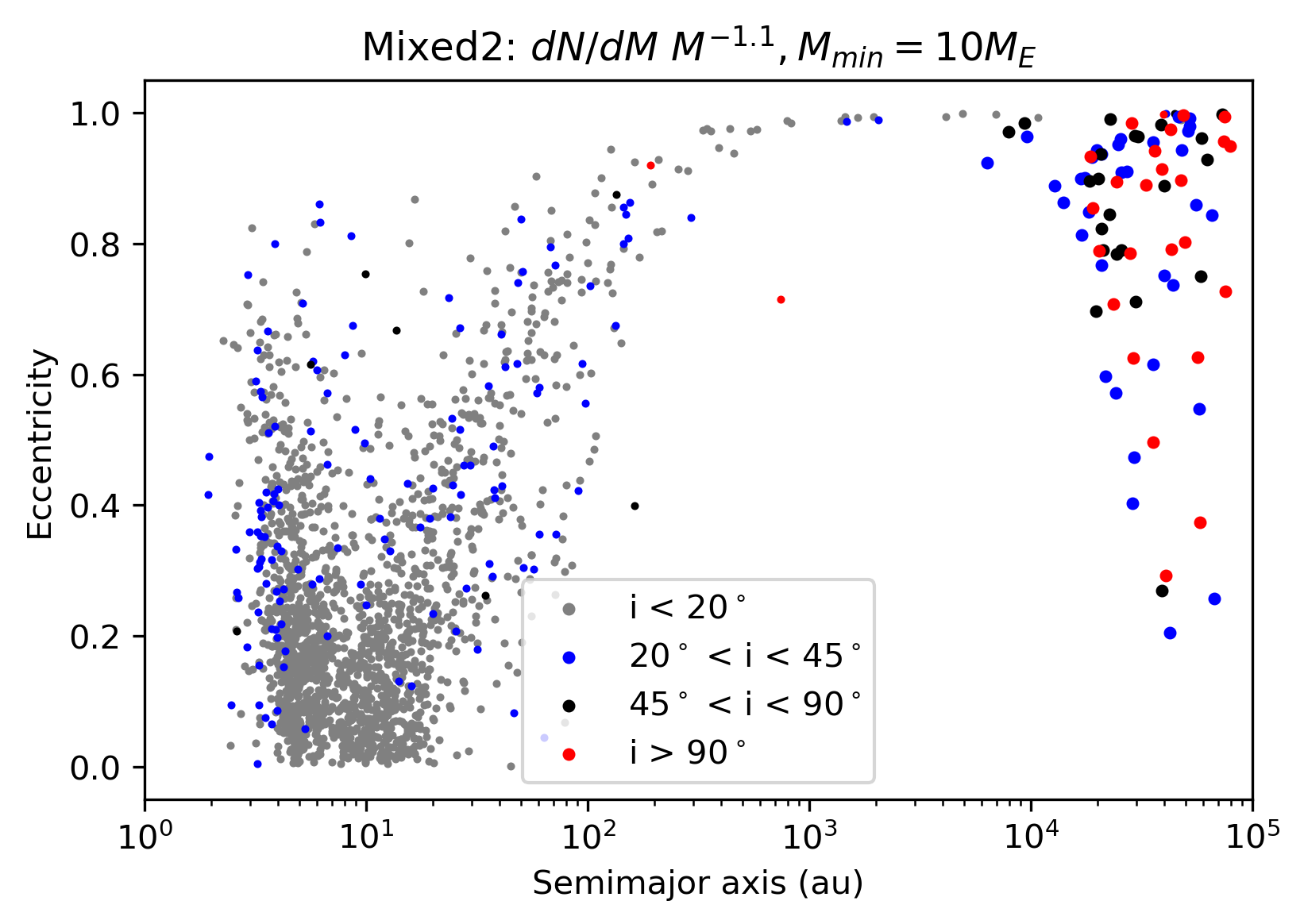}
\\
\includegraphics[width=\columnwidth]{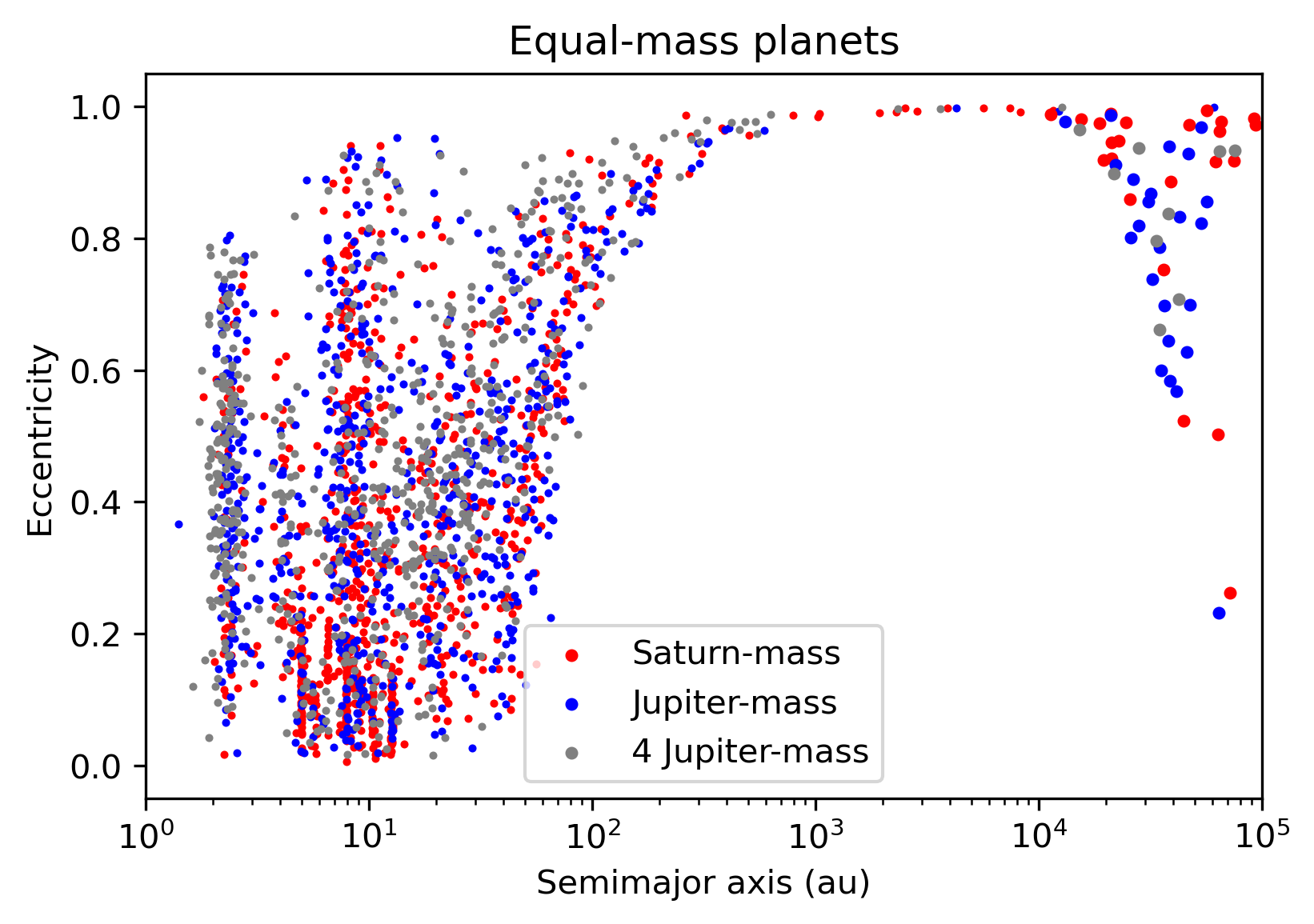}
\includegraphics[width=\columnwidth]{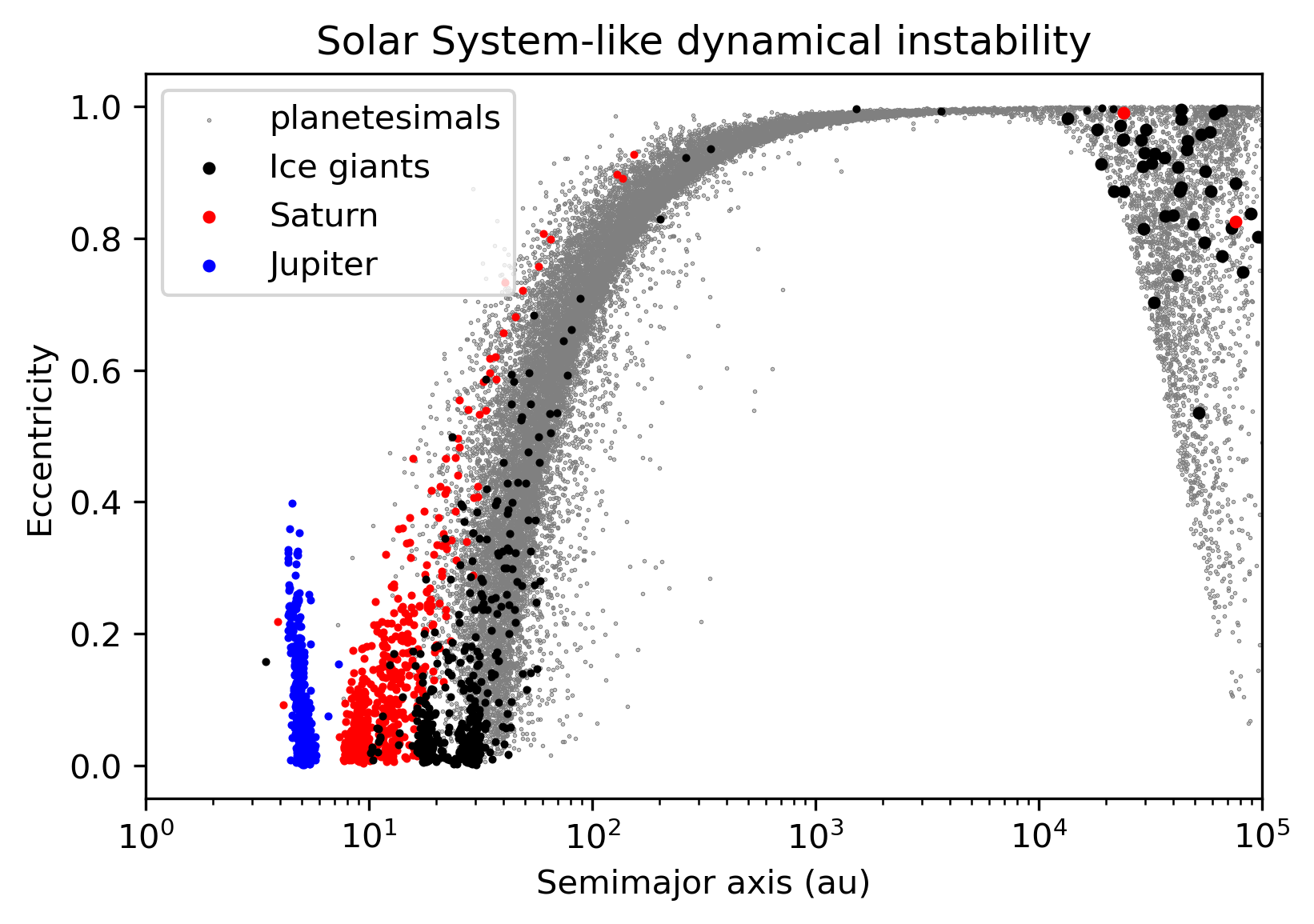}
    \caption{Final distributions of the planets' orbits in our fiducial sets of simulations. In the {\em mixed1} and {\em mixed2} sets (top panels), the color corresponds to the final inclination of each planet's orbit with respect to the starting plane.  In the `Equal-mass' planets set (bottom left), the color corresponds to the planet mass; please note that in each simulation all three planets were the same mass. In the `Solar System-like' set, the color corresponds to the planet, and the small grey dots represent planetesimal particles.  In all panels, Oort cloud planets particles are larger. }
    \label{fig:aei}
\end{figure*}

Figure~\ref{fig:aei} shows the final orbital distribution of the planets in all of our fiducial sets of simulations. Each set shows the same general pattern, with three key features. First, planets inside 10-20 au have a broad distribution of orbital eccentricities.  These are the survivors of planet-planet scattering that provide analogs to the observed giant exoplanets, and our simulations adequately reproduce the observed eccentricity distribution~\citep{raymond10}. Second, there is a long tail of scattered planets that extends from the inner region out to ever-larger semimajor axes, with eccentricities that asymptotically approach unity.  These are the planets that are in the process of being ejected, as they continue to undergo close approaches when they cross the orbits of other planets at periastron.  Third, there is a tail of planets with semimajor axes larger than $\sim 10^4$~au with eccentricities below 1.  These are Oort cloud planets; they followed the ejection track with eccentricities approaching 1, but torques from the Galactic tide decreased their eccentricities sufficiently to lift their periastron distances outside of the realm of surviving planets (as in the simulation from Fig.~\ref{fig:example}).  This is analogous to how the Sun's Oort cloud was populated~\citep{heisler86,duncan87,fernandez97,brasser13b,dones15}.  

The inclinations of surviving planets at Jupiter- to Saturn-like distances tend to have modest inclinations, typically below $20^\circ$ but with a tail extending up to $\sim 45^\circ$ (see top two panels in Fig.~\ref{fig:aei}).  Planets scattered out to very wide orbits on the pathway to ejection tend to maintain similar, modest inclinations.  However, planets captured into their stars' Oort clouds have very broad inclination distributions, with a large fraction of planets on retrograde orbits.  This is, again, similar to the case of Oort cloud comets, whose inclinations are isotropically distributed with respect to the ecliptic~\citep[e.g.][]{dones15}.  

We define Oort cloud planets in our simulations as those with semimajor axes larger than $10^3$~au and periastron distances larger than $10^2$~au. Our simulations show a trend toward a higher rate of planets trapped in the Oort cloud for lower-mass planetary systems.  Among simulations with equal-mass planets there was a rough mass trend, with the most massive planets producing fewer Oort cloud planets: planets were trapped in the Oort cloud in 4.6\%, 4.8\%, and 1.8\% of simulations with Saturn-mass, Jupiter-mass, and 4 Jupiter-mass planets, respectively.  This trend is likely due to the fact that higher-mass planets undergo a smaller number of (stronger) gravitational encounters before ejection~\citep{raymond10}, thus producing a random walk in orbital energy with larger steps than for lower-mass planets. The fraction of simulations that produced a planet in the Oort cloud was 4\% for the {\em mixed1} set and 8.1\% for the {\em mixed2} simulations.  

The most efficient population of the Oort cloud appears to happen when lower-mass ($\sim$Neptune-mass) planets are scattered by Saturn- to Jupiter-mass planets. This outcome is consistent with previous models of Oort cloud formation that varied the masses of the Sun's planets~\citep{lewis13}, and also with trapping of scattered planets in cluster environments~\citep{izidoro23}. Solar System-like instabilities produced the highest overall rate of Oort cloud planets, with 9\% of simulations trapping a planet in the Oort cloud. This includes two systems in which Saturn was scattered by Jupiter and ended up trapped in the Oort cloud.  Such an outcome is not consistent with Solar System's dynamical evolution.  If we restrict ourselves to systems that are broadly consistent with the Solar System's present-day architecture (by requiring a final eccentricity lower than 0.1 for Jupiter), an ice giant was trapped in the Oort cloud in 7\% of simulations.  

Oort cloud planets tend to be lower in mass than their surviving counterparts.  When dynamical instabilities take place in systems with mixed planet masses, lower-mass planets are more commonly ejected -- and when they survive, they tend to have higher eccentricities~\citep{juric08,raymond10}.  The median mass of surviving planets interior to 10~au in the {\em mixed1} set is $1.76 M_J$, but the median mass of Oort cloud planets in those simulations is $0.63 M_J$.  In the {\em mixed2} set this difference is even more striking, as the median mass of surviving inner planets is $0.9 M_J$ but the median mass of Oort cloud planets is $0.092 M_J \approx 30 \mearth$.  Given that the exoplanet occurrence rate increases sharply for lower-mass planets, including past the snow line~\citep{suzuki16b}, we expect the population of Oort cloud planets to be dominated by low-mass planets.  We return to this issue in Section 3.1.

\subsection{Effect of the local Galactic density}

\begin{figure}
\includegraphics[width=\columnwidth]{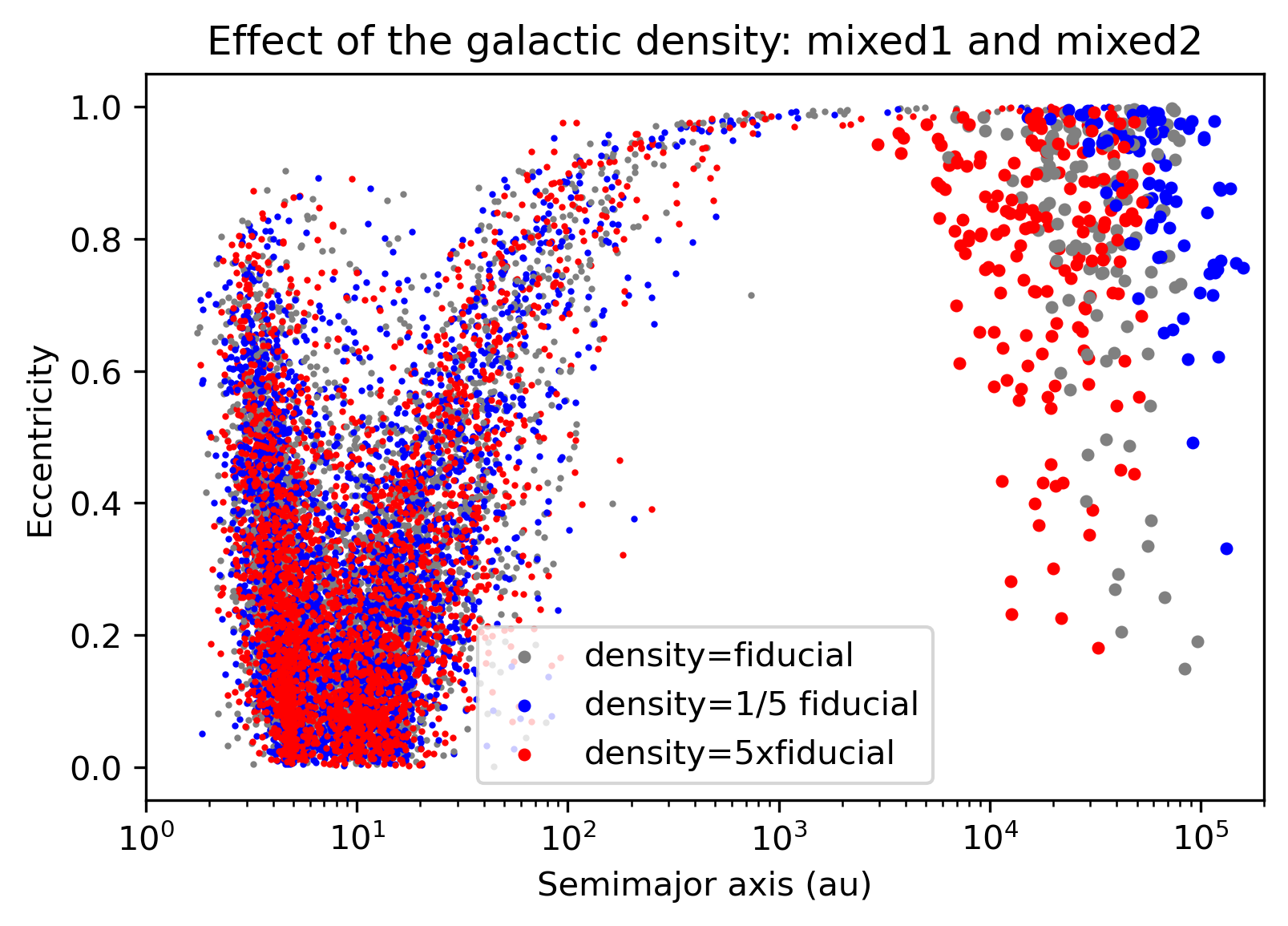}
    \caption{Final distribution of wide-orbit planets in the {\em mixed1} and {\em mixed2} sets of simulations, including those with higher or lower galactic densities.  It is clear that a higher galactic density correlates with trapping scattered planets on closer orbits.}
    \label{fig:galdens}
\end{figure}

The local Galactic density is a central key factor controlling the capture of planets in the Oort cloud.  A star's tidal radius $R_{tide}$ -- beyond which an object is no longer bound to its star -- can be written as~\citep{tremaine93,wyatt17}:
\begin{equation}
    R_{tide} \approx 1.7 \times 10^5 \mathrm{au} \, \left(\frac{M_\star}{M_\odot}\right)^{1/3} \rho_0^{-1/3},
\end{equation}
\noindent where $M_\star$ is the stellar mass.  We performed 1000 additional simulations of the {\em mixed1} and {\em mixed2} sets, varying the local Galactic density $\rho_0$ by a factor of 5 and 1/5.  This corresponds to a radial shift of 1.6 scale lengths~\citep[the radial scale length of the Galaxy is debated, with estimates ranging from 1 to 5 kpc; see][and references therein]{amores17}.  For an increase in $\rho_0$ by a factor of 5, $R_{tide} = 99,400$~au and for a decrease in $\rho_0$ by a factor of 5, $R_{tide} \approx 290,000$~au. We set these values as ejection radii in running additional simulations, while simultaneously increasing the strength of the galactic torque acting on the planets (see Eq. 1).  

For a higher Galactic density, planets are trapped on closer orbits and at a higher rate than for lower Galactic densities.  Figure~\ref{fig:galdens} shows the outcome of our planet-planet scattering simulations for different values of $\rho_0$. For $\rho_0$ of 1/5, 1, and 5 times the fiducial value, the median orbital radius of Oort cloud planets were 60700~au, 34300~au, and 18700~au, respectively.  A planet was trapped in the Oort cloud in 4.2\% of simulations with $\rho_0 = \rho_{fid}/5$, in 6.1\% of simulations with $\rho_0 = \rho_{fid}$, and 8.5\% of simulations with $\rho_0 = 5 \rho_{fid}$. These values include both the {\em mixed1} and {\em mixed2} sets of simulations; the single highest trapping efficiency was for the {\em mixed2} set with $\rho_0 = 5 \rho_{fid}$, in which 9.8\% of simulations produced an Oort cloud planet.

These trends are a consequence of the Galactic tide.  For a higher local Galactic density, the Galactic tidal force can influence planets' orbits closer to their host stars and therefore lift their perihelia and trap them on smaller orbital radii.  While the tidal radius is smaller for a high Galactic density, the stronger tidal force traps a larger fraction of planets than in lower Galactic density.  However, while we do not model their long-term evolution, we suspect that planets formed in higher Galactic densities may be shorter lived because they will undergo faster periastron cycling and be subjected to stronger stellar encounters.  This would go in the direction of balancing the long-term frequencies of Oort cloud planets across the Galaxy.

\section{Discussion}

\subsection{Comparison with previous work}
\cite{portegies21} simulated the scattering of planetesimals and planets orbiting stars with a range of masses on different orbits within the Galaxy, and with corresponding temporal variations of the tidal radius.  Their planetary systems were similar to our equal-mass sets of simulations, as they included planets with near-equal masses (with an `oligarchic' layout), although their systems were at 10-100~au.  \cite{portegies21} found a typical capture rate of Oort cloud planets of $\sim 2$\%.  While this is lower than that in most of our sets of simulations, it is a decent match to the set with equal-mass planets of $4 M_J$, which is close to the median planet mass in \cite{portegies21}.  We interpret this as confirmation that our results are broadly realistic despite our simplified implementation of a fixed tidal radius.

\subsection{Long-term survival and dynamics of Oort cloud planets}

The population of Oort cloud planets only declines very slowly in time. Our fiducial runs simulated the main phase of capture of planets within the Oort cloud, but not their long-term survival. We addressed this question in two steps.  We first ran all {\em mixed1} and {\em mixed2} simulations to 500 Myr. The {\em mixed1} set contained 40 Oort cloud planets at 100 Myr; it lost 11 (27.5\%) of its Oort cloud planets between 100 and 500 Myr, but gained 10 new ones.  The {\em mixed2} set contained 81 Oort cloud planets at 200 Myr; it lost 13 (16\%) between 200 and 500 Myr, but gained 8 new ones.  Next, we selected all of the {\em mixed1} and {\em mixed2} simulations with Oort cloud planets at 500 Myr and ran them out to 1 Gyr.  Combining both sets, 81 out of 114 Oort cloud planets (71\%) survived.  This is an underestimate because these simulations did not allow for the capture of planets in the process of being ejected (from what \cite{portegies21} calls the 'Oort cloud conveyor belt').

In our long-term simulations, Oort cloud planets were lost mainly because the same torquing that decreases the eccentricities of wide-orbit planets (to trap them in the Oort cloud) acts to make their eccentricities increase again, bringing them back into dynamical contact with the inner planetary system~\citep[as for wide binary stars; ][]{kaib13}. This happened to the Oort cloud planet from Fig.~\ref{fig:example} after $\sim 600$~Myr; while the planet's orbit remained within the Oort cloud, it received a visible energy kick and corresponding bump in its semimajor axis.  The timescale for this torquing depends mainly on the planet's orbital radius and inclination, and is typically between 0.1 and several Gyr. This second- (or third-) generation scattering can lead to the planet's ejection.  In nature, this torquing is unlikely to be as easily repeated as in our simulations, as the phases of scattered planets can be randomized  (e.g. by passing stars and the radial component of the galactic torque) such that torques do not exactly reverse the planet's eccentricity evolution.  Of course, passing stars can strip loosely bound Oort cloud planets~\citep[e.g.][]{matese92} and transform them into free-floating planets like the ones observed via microlensing~\citep{mroz17} and direct imaging~\citep{miretroig22}.

\subsection{Frequency of Oort cloud planets}

To estimate the fraction of stars that may host Oort cloud planets requires two quantities: the occurrence rate of gas giants and the Oort cloud trapping efficiency.

Radial velocity and microlensing surveys find that $\sim 10-20\%$ of Sun-like stars host gas giants~\citep[e.g][]{mayor11,suzuki16b,clanton16}, most located at 1-3~au or beyond~\citep{mayor11,fernandes19,lagrange23}.  The occurrence rate of gas giants increases with stellar metallicity~\citep{fischer05}, especially among systems showing strong signs of planet-planet interaction~\citep{dawson13}.  The occurrence rate also increases as a strong function of the stellar mass~\citep{johnson07}.  

The trapping frequencies for our different sets of simulations ranged from 1-10\%.  The simulations that most naturally reproduce the exoplanet eccentricity distribution -- the {\em mixed1} and {\em mixed2} sets -- trapped planets in the Oort cloud in 4-8\% of simulations, although matching the observed mass-eccentricity correlation also requires a contribution from equal-mass systems~\citep{raymond10}, which had trapping efficiencies of 1-5\%.  

Given the uncertainties in occurrence rates, we assume that 1-10\% of all stars host gas giants, integrated over all spectral types.  Giant planet systems have a mixture of orbital architectures, and our results suggest that roughly 5\% of unstable systems will trap a planet in the Oort cloud for the Sun's local density.  To match the observed eccentricity distribution, 75-95\% of all giant planet systems must be the survivors of instability~\citep{juric08,raymond10}.  Put together, this implies an Oort cloud planet frequency between $3.75 \times 10^{-4}$ and $4.75 \times 10^{-3}$. This amounts to one Oort cloud planet for every $\sim$210-2700 stars.  This value only applies to planets within the mass range of our simulations.  In real systems, planetary embryos and lower-mass planets are frequently ejected, and may outnumber the ejected planets by an order of magnitude or more~\citep{barclay17}. The rate of Oort cloud Mars-mass or Earth-mass planets may thus be far higher than estimated above. 

Our simulations apply to instabilities that take place after the dispersal of the stellar birth cluster, a few to ten Myr after the start of star and planet formation~\citep[e.g.][]{adams10}. Instabilities during the cluster phase can trap planets on orbits with radii of hundreds to thousands of au~\citep{izidoro23}, but the high local density makes it impossible to reach $10^{4-5}$~au simply because those lie beyond the tidal radius~\citep[see Eq. 2;][]{tremaine93,wyatt17}. While the timing of giant planet instabilities is hard to constrain directly, the existence of a rich population of free-floating planets in the Upper Scorpius star-forming association hints that many instabilities may indeed happen early~\citep{miretroig22}.  Our calculation may therefore overestimate the true abundance of Oort cloud planets. Nonetheless, there is circumstantial evidence that some systems undergo late dynamical instabilities -- such as the abundant dust produced in the 1 Gyr-old $\eta$~Corvi system~\citep[e.g.][]{lisse12} -- so it appears unavoidable that Oort cloud planets do exist.

\section*{Acknowledgements}
We thank the anonymous referee for a constructive report.  S.N.R. thanks the CNRS's PNP and MITI programs for support.  A.I. acknowledges support from NASA grant 80NSSC18K0828 and the Welch Foundation grant No. C-2035-20200401. N.A.K.'s contributions were supported by NASA Exoplanets Research Program grant 80NSSC19K0445 and NSF CAREER Award 1846388.


\section*{Data Availability}

All simulations and analysis code in this paper will be made available upon reasonable request.









\bsp	
\label{lastpage}
\end{document}